\documentclass[prb,showpacs,twocolumn,amsmath,amssymb,floatfix]{revtex4}

\usepackage[utf8]{inputenc}
\usepackage{graphicx}
\usepackage{color}
\usepackage{bm}

\graphicspath{{Figures/}{Figures_old/}}

\newcommand{\pdagger}{{\phantom{\dagger}}}
\newcommand{\dt}{\Delta\tau}
\newcommand{\reff}[1]{Fig.\ \ref{fig:#1}}
\newcommand{\reffa}[1]{Fig.\ \ref{fig:#1}(a)}
\newcommand{\reffb}[1]{Fig.\ \ref{fig:#1}(b)}
\newcommand{\reffc}[1]{Fig.\ \ref{fig:#1}(c)}

\newcommand{\refq}[1]{(\ref{eq:#1})}
\newcommand{\mymin}{{\text{min}}}
\newcommand{\refs}[1]{Sec.\ \ref{sec:#1}}

\begin{document}

	\title{Mott transitions in the half-filled SU($\bm{2M}$) symmetric Hubbard model}

	\author{N.~Bl\"umer}
	\email{Nils.Bluemer@uni-mainz.de}
\author{E.~V.~Gorelik}
\affiliation{Institute of Physics, Johannes Gutenberg University, 55099 Mainz, Germany}

\date{\today}

  \begin{abstract}
	The Hubbard model with large orbital degeneracy has recently gained relevance in the context of ultracold earth alkali-like atoms. We compute its static properties in the SU($2M$) symmetric limit for up to $M=8$ bands at half filling within dynamical mean-field theory, using the numerically exact multigrid Hirsch-Fye quantum Monte Carlo approach. Based on these unbiased data, we establish scaling laws which predict the phase boundaries of the paramagnetic Mott metal-insulator transition at arbitrary orbital degeneracy M with high accuracy.		
  \end{abstract}
  \pacs{67.85.-d, 03.75.Ss, 71.10.Fd, 71.30.+h}
  \maketitle


\section{Introduction}\label{sec:intro}
  The interaction-induced Mott transition between a metal and a paramagnetic insulator is
	central to the field of strongly correlated electron systems.\cite{Imada1998} Much insight
	into this phenomenon has been gained in numerical studies of the
	single-band Hubbard model within dynamical mean-field theory (DMFT).\cite{Georges96}
	In particular, the phase diagram and the behavior of characteristic observables
	(such as the effective mass) have been established with high
	precision\cite{Schlipf1999,Rozenberg1999,Krauth2000,Kotliar2000,Joo2001,Bulla2001,Tong2001,Kotliar2004,Bluemer05ab}
	 - despite the lack of analytic solutions. 
	
	While the single-band assumption is rather crude in correlated solids 
    (see below),
	it can be quite accurate for two-flavor mixtures of ultracold fermions on optical lattices. Since, in addition, the effective interaction between neutral ultracold alkali atoms (in their electronic ground state) is very short-ranged, such systems appear as nearly perfect finite-size realizations of the single-band Hubbard model, with the prospect of addressing some of the open questions (e.g. regarding high-$T_c$ superconductivity) via the tunable {\it quantum simulation} of the underlying Hamiltonian. An important step in this direction was the recent experimental verification of the Mott transition in cold-atom systems,\cite{Joerdens2008,Schneider2008} for which accurate quantitative predictions based on DMFT 
	were essential.
	
	The low-energy electronic properties of correlated solids are usually determined by $d$ orbitals, which are five-fold degenerate (per spin) in the atomic limit. This degeneracy is partially lifted by the crystal field, resulting, e.g., in a three-fold degenerate $t_{2g}$ band and a two-fold degenerate $e_g$ band for cubic symmetry. Each of these bands is characterized by a local potential plus various Hund rule couplings (which can also couple inequivalent bands). Thus, the multi-orbital case is not only richer physically\cite{Kotliar1996,Han1998,Koga2002,Ono2003,Pruschke2005} (including the possibility of orbital-selective Mott transitions\cite{Knecht2005,Jakobi2009,vDongen2006,Anisimov2002,Koga2005}), but is complex already by the number of parameters. In addition, obtaining accurate numerical results rapidly becomes costly and challenging with increasing number $M$ of orbitals. In fact, some methods such as the numerical renormalization group (NRG) become impractical beyond $M=2$ orbitals. As a consequence, few properties of the multi-orbital Hubbard model can be considered well-established with high precision,
even at the DMFT level.

	However, there is a unique generalization of the single-band Hubbard model to arbitrary degeneracy 
	which avoids the introduction of 
	additional parameters: In the SU($2M$) symmetric Hubbard model, all spins and orbitals are equivalent, i.e., share the same local potential, the same hopping matrix elements, and they are coupled by the same local interaction. In other words, the phase space of this particular multi-orbital model is identical to the single-band case. Moreover, interesting analytic insights have been obtained in the limit of large band multiplicity $M\to\infty$, including an exact expression for the critical interaction of the ground-state metal-insulator transition (at half band filling) as well as scaling arguments for the finite-temperature critical end point.\cite{Florens_PRB02b} Thus, the sequence of models obtained by varying $M$ connects two well-established -- and somewhat special -- limits ($M=1$, $M=\infty$), while the intermediate regime $M=2,3,\dots$ shares many characteristics with generic multi-orbital models, including numerical difficulties. Indeed, the SU($2M$) Hubbard model has, so far, been explored in this regime only using approximate methods, namely the dynamical slave-rotor approximation (DSR),\cite{Florens_PRB02a} the projective self-consistent method (PSCM),\cite{Moeller1995,Kotliar1996} and the self-energy functional approximation (SFA) with one bath site per orbital.\cite{Inaba_PRB05} A fully controlled treatment is clearly desirable on fundamental grounds and as a solid starting point for generic multi-orbital physics.
	
	Quite recently, the SU($N$) Hubbard model (with total degeneracy $N>2$) has also become of direct physical relevance, namely in the ultracold atom context: In rare-earth atoms, a large number of internal states can be addressed, which are essentially decoupled from the valence electrons. Consequently, all atoms in the electronic ground state experience the same optical potential and have the same pairwise interactions;\cite{Fukuhara2007,DeSalvo2010} a mixture with N internal states on an optical lattice can, therefore, realize the SU($N$) symmetric Hubbard model. A Mott insulating state has already been observed in a SU(6) symmetric system of fermionic ytterbium atoms ($^{173}$Yb) on a cubic optical lattice,\cite{Taie2012} opening the door to detailed experimental investigations of Mott metal-insulator transitions in SU($N$) symmetric Hubbard models (with $N>2$).
This breakthrough has sparked theoretical interest in both SU($N$) Hubbard\cite{Cai2012ab,Tokuno2012,Bonnes2012} and Heisenberg\cite{Messio2012} systems, with initial studies being limited to one spatial dimension\cite{Bonnes2012,Messio2012} and to a slave-particle method,\cite{Tokuno2012} respectively. 

	In this work, we construct the phase boundaries	of the Mott transition at half filling and for up to
	$M = 8$ bands, based on numerically exact multigrid Hirsch-Fye quantum Monte
	Carlo\cite{Bluemer_multigrid,Gorelik2009} estimates of characteristic observables.
	We also derive scaling laws which predict the phase boundaries for arbitrary orbital degeneracy $1\le M\le \infty$ with high accuracy.

    In \refs{models}, we establish our notation and relate the SU($N$) symmetric Hubbard model to generic multi-band models. We also introduce the DMFT in the present context, discuss our choice of lattice type and energy scales, and characterize our DMFT impurity solver. 
In \refs{phase}, we specify the procedure to determine the phase boundaries, briefly summarize literature data for the SU($2M$) symmetric Hubbard model, and present numerically exact results for $M=2$, $4$, and $8$. 
Based on these data, we deduce in \refs{scaling} the universal scaling of the critical parameters with spin and orbital degeneracy and establish the collapse of finite-$M$ data onto an universal phase diagram.
	

\section{Model and methods}\label{sec:models}

The general Hubbard model for $M$ equivalent electronic orbitals (e.g., $M=3$ t$_{2g}$ orbitals) with nearest-neighbor hopping $t$ and SU(2) invariant Hund's rule coupling $J$ has the form:
\begin{eqnarray}
 H \! &=&\! \sum_{m=1}^M \bigg[ -t \sum_{\langle ij\rangle \sigma}\! \big(c_{im\sigma}^\dagger  c_{jm\sigma}^\pdagger + \text{h.c.}\big)
+\, U\sum_i n_{im\uparrow}^{\phantom{\dagger}} n_{im\downarrow}^{\phantom{\dagger}} \bigg]
\nonumber \\
&&+\, \tfrac{1}{2}\sum_{m\not=m'} \bigg[\, \sum_{i\sigma\sigma'}\big(U' -\delta_{\sigma\sigma'}^{\phantom{\dagger}}J\big)\, n_{im\sigma}^{\phantom{\dagger}} n_{im'\sigma'}^{\phantom{\dagger}} \nonumber\\
 &&+\, J\sum_{i\sigma}  c_{im\sigma}^\dagger \Big(  c_{im'\bar{\sigma}}^\dagger  c_{im\bar{\sigma}}^{\phantom{\dagger}} + c_{im\bar{\sigma}}^\dagger  c_{im'\bar{\sigma}}^{\phantom{\dagger}} \Big)  c_{im'\sigma}^{\phantom{\dagger}}\bigg]\label{eq:J}
\end{eqnarray}
Here, the first line can be viewed as $M$ versions of the regular 1-band Hubbard model with (intraorbital) on-site Hubbard interaction $U$; $\langle ij\rangle$ denotes pairs of nearest-neighbor sites $i$ and $j$, $\sigma\in\{\uparrow, \downarrow\}$ the spin. The coupling between these orbitals is provided, in general, by the interorbital density-density interaction $U'$ and the Hund's rule coupling $J$; the latter contains both an Ising-type contribution, coupling to the spin densities $n_{im\sigma}^{\phantom{\dagger}} \equiv c_{im\sigma}^\dagger c_{im\sigma}^\pdagger$ (second line), as well as spin-flip and pair-hopping terms (third line). In the limit $J\to 0$, the relation $U=U'-2J$ implies that the inter- and intraorbital Hubbard interaction become equal: $U'\to U$. Thus, at $J=0$, spin and orbitals are fully equivalent and one arrives at the SU($N$) symmetric Hubbard model with $N=2M$ even:
\begin{equation}
 H = -t \sum_{\alpha=1}^{N} \sum_{\langle ij\rangle} \big(c_{i\alpha}^\dagger  c_{j\alpha}^\pdagger + \text{h.c.}\big)
  + U \sum_{\alpha<\alpha'} n_{i\alpha} n_{i\alpha'}\,,\label{eq:SUN}
\end{equation}
where $\alpha$ is a combined spin-orbital index. This is precisely the situation which has been realized, within the single-band approximation and up to the confining potential, using rare-earth (earth alkali-like) ytterbium atoms ($^{173}$Yb) on a simple cubic optical lattice.\cite{Taie2012} Note that exchange terms as appearing in \refq{J} at $J\not=0$ require a unique classification of the internal degree of freedom $\alpha\in\{1,N\}$ in terms of the ``spin'' variable  $\sigma\in\{\uparrow, \downarrow\}$ and cannot arise in the SU(N) symmetric case, where all values of $\alpha$ are fully equivalent.

Paramagnetic Mott metal-insulator transitions (MITs) can be expected for this model at all integer fillings $n\equiv \sum_\alpha\langle n_{i\alpha}\rangle\in \{1,2,\dots, N-1\}$ (while $n=0$, $n=N$ correspond to band insulators).
For $N=2M$ being even (as always in the electronic context) this includes the case of half filling $n=N/2$, where one then expects the largest critical interaction (compared to pairwise equivalent MITs at fillings $n=N/2\pm 1$, $n=N/2\pm 2$, \dots).\cite{Rozenberg1997}
In the cases of odd $N\ge 3$, not to be considered in this paper, 
the Mott plateaus at fillings $n=N/2\pm 1/2$ are separated by a metallic phase with unique ``semi-compressible'' properties.\cite{Gorelik2009} 

The DMFT reduces the lattice problem \refq{SUN} to a single-impurity problem,\cite{Georges96} with the same local SU($N$) invariant interaction terms, which has to be solved self-consistently.\cite{Rozenberg1997} For homogeneous phases, the lattice properties enter only via the corresponding tight-binding density of states $\rho(\varepsilon)$. In line with previous studies, we choose the semi-elliptic form associated with the Bethe lattice\cite{Kollar2005} and set the energy scale as $t\sqrt{Z}=1$ (for coordination number $Z$), which implies unit variance of the density of states: $\int_{-\infty}^\infty d\varepsilon \varepsilon^2 \rho(\varepsilon)=1$. Our numerical results can be translated, e.g., to the cubic lattice (in units of the hopping $t$) by multiplying interactions, energies, and temperatures by $\sqrt{Z}=\sqrt{6}\approx 2.45$.\cite{fn:scales}

 As the interaction couples only to spin-orbital densities [i.e., spin flip and pair hopping terms, as arising in the general model \refq{J} for $J\not=0$, are absent], DMFT solutions can be obtained using quantum Monte Carlo (QMC) impurity solvers without any sign problem for arbitrary density. The Hirsch-Fye algorithm\cite{Hirsch86,Bluemer2007} discretizes the imaginary-time path integral expression for the Green function into $\Lambda$ time slices of uniform width $\dt=\beta/\Lambda$, where $\beta=1/T$ (for $k_{\text{B}}\equiv 1$); a Hubbard-Stratonovich transformation replaces the electron-electron interaction (for each pair $\alpha<\alpha'$) at each time step by a binary auxiliary field which is sampled using standard Markov Monte Carlo techniques. 
In this work, we use a multi-grid implementation\cite{Bluemer_multigrid,Gorelik2009} and, thereby, demonstrate that its inherent elimination of Trotter errors from the Green function and from observables works reliably and accurately even for a large number $M$ of bands and $M (2M-1)$ Hubbard-Stratonovich fields. Consequently, our results are free of significant systematic bias, i.e., exact within statistical error bars.
These statistical errors are reduced, compared to a generic $M$-band model, by employing the SU($2M$) symmetry, i.e., by averaging Green functions and related observables over all $2M$ values of the internal degree of freedom $\alpha$ and the double (or pair) occupancy over all $M(2M-1)$ pairs $\alpha<\alpha'$.


\section{Determination of MIT phase boundaries}\label{sec:phase}

It the noninteracting limit $U\to 0$, the Hamiltonian \refq{SUN} reduces to the corresponding tight-binding model; due to the degeneracy, the system is then metallic at all densities $0<n<N$. In contrast, the energy levels become discrete in the atomic limit $t\to 0$; at integer filling, the system is then an insulator. The question of how the evolution between these two limits takes place, e.g. as a function of varying $U$ at constant $t$, has been a matter of debate for a long time.\cite{Schlipf1999, Mott1968, Brinkmann1970, Imada1998} It is now well-established  that the (paramagnetic) metallic and paramagnetic insulating phases are separated, at low temperatures and within DMFT, by a sharp transition line in the single-band case (i.e., for $N=2M=2$).

This transition is of first order at temperatures $0<T<T^*$ (thick blue line in the inset of \reff{SFA_DSR}), evolving to  second order both at the critical end point ($T^*$, $U^*$) and in the limit $T\to 0$ (and $U\to U_{c2}$); here and in the following, we use the notation $U_{c2}=U_{c2}(T\!=\!0)$ and $U_{c1}=U_{c1}(T\!=\!0)$ for ground-state values. Due to its mean-field character, the DMFT self-consistency equations do not directly yield the critical line $U_c(T)$; instead, one finds, at $T<T^*$, coexistence of metallic and insulating solutions in the range $U_{c1}(T)\le U\le U_{c2}(T)$ (indicated by circles in the inset of \reff{SFA_DSR}). 
The determination of $U_c(T)$ within these boundaries requires a comparison of free energies, which are not directly accessible in QMC based approaches (but can be obtained via integration of thermodynamic relations\cite{BluemerPhD}).

As discussed in \refs{intro}, the situation is expected to be quite similar in the multi-band case $M>1$. Specifically, DMFT should yield a coexistence region of metallic and insulating solutions at arbitrary $M$, including the limit $M\to \infty$. In this limit, $U_{c2}$ was shown\cite{Florens_PRB02b} to approach $4 |E_0|$, where $E_0$ is the noninteracting ground-state (kinetic) energy; for the Bethe lattice, $E_0=-8M/(3\pi)\approx -0.85 M$. 
 However, as the single-band case deviates from this asymptotic value by nearly a factor of two, numerics at $1<M<\infty$ are needed in order to derive quantitative predictions from this analytic result. Linearized DMFT\cite{Ono2003} is not sufficient in this respect; its prediction $U_{c2}=4M+2$ (for arbitrary $M$) is consistent with the exact asymptotic result only regarding the power (in $M$), but the prefactor (4 instead of $32/(3\pi)\approx 3.4$) is obviously incorrect.
Remarkably, the critical interaction at finite temperatures scales differently: $U^* \propto M^{1/2}$, as was argued convincingly analytically.\cite{Koch1999}  In this case, the analytic considerations do not even yield a prefactor; thus, quantitative predictions regarding $U^*$ are completely dependent on accurate numerical results for sufficiently large values of $M$.

\subsection{Previous results for $\bm{M\ge 2}$}

So far, the complete DMFT coexistence regions have been computed at $M>1$ only using the dynamical slave-rotor formalism\cite{Florens_PRB02a} and using the self-energy functional approach.\cite{Inaba_PRB05} The former is an approximate impurity solver which contains a free parameter and had been tested quantitatively only for $M=1$. Consequently, the accuracy of its results at $M>1$ (and even in the limit $M\to\infty$) is, a priori, completely unclear. In contrast, the SFA \cite{Potthoff2003} is based [in the variant used in Ref.\ \onlinecite{Inaba_PRB05}, known as the dynamical impurity approximation (DIA)] on a discretization of the DMFT dynamical bath; it reduces to the DMFT in the limit of an infinite number of bath sites, $N_b\to\infty$. So this method is numerically exact (within DMFT); however, an unknown bias remains for a finite value of $N_b$, in particular for the ``two-site SFA'' with a single bath site (per interacting orbital), as employed in Ref.\ \onlinecite{Inaba_PRB05}. It is, a priori, unclear, how this bias evolves with $M$ (at fixed $N_b/M$).

As shown in \reff{SFA_DSR}, the DSR (dotted lines) and the SFA (dashed lines) both yield coexistence regions for $M=2$ and $M=4$ which have shapes similar to those in the single-orbital case $M=1$. 
\begin{figure}[t] 
  \includegraphics[width=\columnwidth]{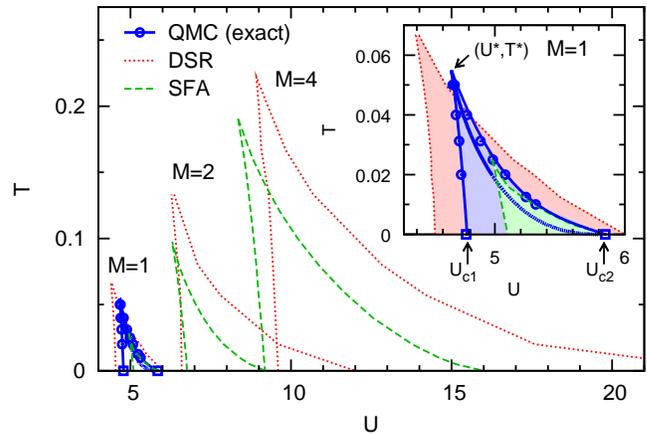}
  \caption{(Color online) Previous results for the Mott metal-insulator transition in the SU($2M$) symmetric Hubbard model within dynamical mean-field theory: coexistence phase diagrams for band degeneracy  $M=1$, 2, and 4 obtained using the dynamical slave-rotor approximation (DSR)\cite{Florens_PRB02a}  and the self-energy functional approximation (SFA),\cite{Inaba_PRB05} respectively, in comparison with numerically exact quantum Monte Carlo (QMC) data for $M=1$. Inset: magnified view for $M=1$.\label{fig:SFA_DSR}}
\end{figure}
At $M=2$, even the critical interactions are in good mutual agreement with a value $U^*\approx 6.3$; however, this agreement seems coincidental, as the DSR estimate of $U^*$ is significantly below (above) the SFA estimate at $M=1$ ($M=4$).
In general, the DSR appears to yield much larger coexistence regions than the SFA. As the DSR is an uncontrolled and comparatively cheap approximation, one might be tempted to put more trust in the SFA results. However, both the DSR and the SFA deviate very significantly from the exact QMC results previously established for $M=1$: as seen in the inset of \reff{SFA_DSR}, the SFA underestimates the critical temperature $T^*$ by about a factor of 2 and the area
of the coexistence region by even more, while the DSR overestimates the latter by nearly the same factor. Given these discrepancies, it is clear that the (previously published) data shown in \reff{SFA_DSR} are not sufficient for verifying the scaling laws discussed above and for determining their prefactors and corrections at finite $M$.

\subsection{Insights from the single-band case ($\bm{M=1}$)}
In order to achieve this goal, we will, in the remainder of this section, determine unbiased coexistence phase boundaries at $M=2$ and $M=4$ and determine $T^*$ and $U^*$ at $M=8$, based on exact QMC data. For completeness and for better illustration of the asymptotic behavior of the relevant observables in the limit $T\to 0$, we will first discuss results for the single-band case ($M=1$), depicted in \reff{Z_D_E_M1}.
\begin{figure}[t] 
  \unitlength0.1\columnwidth
  \begin{picture}(10,11)
	\put(0,10.6){(a)}
	\put(0,7.1){(b)}
	\put(0,3.6){(c)}
  	\put(0,0){\includegraphics[width=\columnwidth]{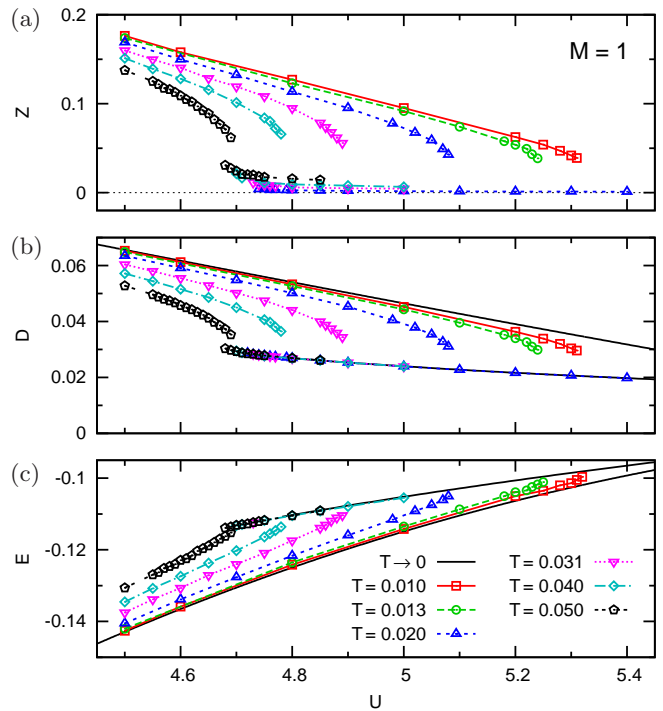}}
  \end{picture}
  \caption{(Color online) Numerically exact DMFT results (symbols), obtained using multigrid HF-QMC, in the vicinity of the Mott metal-insulator transition for $M=1$: (a) quasiparticle weight $Z$, (b) double occupancy (= pair occupancy) $D$, and (c) energy $E$ as a function of the on-site interaction $U$. The extrapolations to the ground state (black solid lines) for $D$ and $E$ include perturbative informations and thermodynamic consistency;\cite{Bluemer05ab} other lines are guides to the eye only. \label{fig:Z_D_E_M1}}
\end{figure}

The quasiparticle weight $Z=m/m^*$ quantifies the renormalization of the quasiparticles in a Fermi liquid by interactions and is closely associated with the inverse linear specific heat: 
$Z(U,T=0) = \gamma(0)/\gamma(U)$ [with energy $E(U,T)=E(U,0)+\gamma(U)\,T^2/2 + {\cal O}(T^4)$]. 
It can be expressed (exactly) in terms of the self-energy $\Sigma(\omega)$ as 
\begin{equation}
  Z^{-1}=1-\partial\, \text{Re} \Sigma(\omega)/\partial \omega\big|_{\omega=0} ; 
\end{equation}
\reffa{Z_D_E_M1} shows corresponding discrete QMC estimates at finite temperatures, based on the value of the self-energy at the first Matsubara frequency $i\omega_1=i\pi T$:
\begin{equation}\label{eq:Z_discr}
  Z^{-1}\approx 1 - \text{Im} \Sigma(i\pi T)/(\pi T) . 
\end{equation}
Clearly, the data set for each temperature (denoted by symbols) is split into two branches: one metallic branch with moderately high values ($Z\gtrsim 0.04$) which extends down to $U=0$ (shown only for $U\ge 4.5$) and an insulating branch where $Z\approx 0$ for the lower temperatures (e.g. $T=0.02$, denoted by upward triangles). Only at the highest temperatures shown ($T=0.04$ and $T=0.05$) do the estimates in the insulating phase reach values up to about 0.01, which is mainly an artifact of the discrete approximation of $Z$, Eq.\ \refq{Z_discr}; in particular, these values of $Z$ have no relation to the specific heat (which is exponentially small in this range).

In contrast, the double occupancy $D=\langle n_{i\uparrow} n_{i\downarrow} \rangle$ (i.e., the probability of a site being occupied by two fermions simultaneously), which is depicted in \reffb{Z_D_E_M1} and related to the interaction energy by $E_{\text{int}}=U D$, has very significant values in both phases. $D$ is independent of temperature at the scale of the figure in the insulating phase, within its temperature-dependent range of stability, i.e., for $U>U_{c1}(T)$. This behavior is expected in a gapped phase, where thermal excitations are suppressed exponentially. It allows high-precision estimates of the ground-state function $D_{\text{ins}}(U)$ from QMC, in particular by extrapolation of high-order coefficients of strong-coupling perturbation theory.\cite{Bluemer05ab} 
On the metallic side, $D$ depends strongly on temperature, especially in the range $T\lesssim T^*\approx 0.055$.\cite{BluemerPhD} As a function of $U$, the shapes of the curves look remarkably similar for $D$ and $Z$; for both observables, the (negative) curvature becomes much stronger near the boundaries of the metallic phase, i.e., at $U\lesssim U_{c2}(T)$.

In comparison, the results for the energy $E=\langle H\rangle$ look nearly linear in \reffc{Z_D_E_M1} as function of $U$ in the same parameter ranges (which implies that the kinetic energy, not shown, has a positive curvature which nearly cancels that of $D$); the values also approach those of the insulating solution (again with invisible temperature dependence) much more closely. As the relation $D=\partial F/\partial U$ for free energy $F=E-TS$ reduces to $D_{\text{met}}(U)=d E_{\text{met}}(U)/dU$ in the ground state, $D(U,T)$ and $E(U,T)$ are not independent at low $T$; this connection as well as the relation between $\gamma$ and $Z$ have made it possible to determine the ground-state energetics [black solid lines in \reffb{Z_D_E_M1} and \ref{fig:Z_D_E_M1}(c)] in the metallic phase as well.\cite{Bluemer05ab}

The crucial point for determining the phase boundaries $U_{c1}(T)$, $U_{c2}(T)$ via data sets such as depicted in each of the panels in \reff{Z_D_E_M1} is that all included data points actually denote converged solutions, i.e., they correspond to fixed points of the DMFT self-consistency cycle, whereas no metallic solutions exist at $U>U_{c2}(T)$ and no insulating solutions exist at $U<U_{c1}(T)$, respectively. Both the verification and the exclusion of such fixed points are very difficult to achieve reliably, as numerical noise (associated with Monte Carlo importance sampling for a finite number of sweeps), systematic bias (e.g. resulting from Trotter errors) and critical slowing down (for $T\approx T^*$ and $U\approx U^*$) can easily lead to false positives or negatives. For this reason, it is essential to monitor several observables at the same time, as deviations from the expected systematics can help to identify artifacts of incomplete convergence or divergence after a (necessarily) finite number of DMFT iterations.

In this manner, we have obtained the phase boundaries shown as circles in the inset of \reff{SFA_DSR} (building upon earlier work\cite{BluemerPhD}) with high precision at finite temperatures $T\ge 0.01$. The squares denote complementary ground-state results for $U_{c1}$ from extrapolated perturbation theory \cite{Bluemer05ab} and for $U_{c2}$ from ED and NRG.\cite{Georges96,Bulla2001} Taken together, these results determine fit functions for the coexistence region (thin solid lines and blue-shaded region) with high precision; we will later test the hypothesis that very similar fits might capture the Mott transition at $M>1$. Note that the (numerically exact) QMC estimate of $U_{c2}(T)$ and its extrapolation to $T\to 0$ agree well with the corresponding SFA estimate (at $T<T^*_{\text{SFA}}$).
The inset of \reff{SFA_DSR} also shows a thick line within the coexistence region that denotes the DMFT estimate of the actual first-order phase transition.\cite{BluemerPhD}

\subsection{QMC results for $\bm{M\ge 2}$}
The quasiparticle weight, as defined above, remains a well-defined and useful concept at arbitrary degeneracy and is shown in \reffa{Z_D_E_M2} for $M=2$. 
\begin{figure}[t] 
  \unitlength0.1\columnwidth
  \begin{picture}(10,11)
	\put(0,10.6){(a)}
	\put(0,7.1){(b)}
	\put(0,3.6){(c)}
  	\put(0,0){\includegraphics[width=\columnwidth]{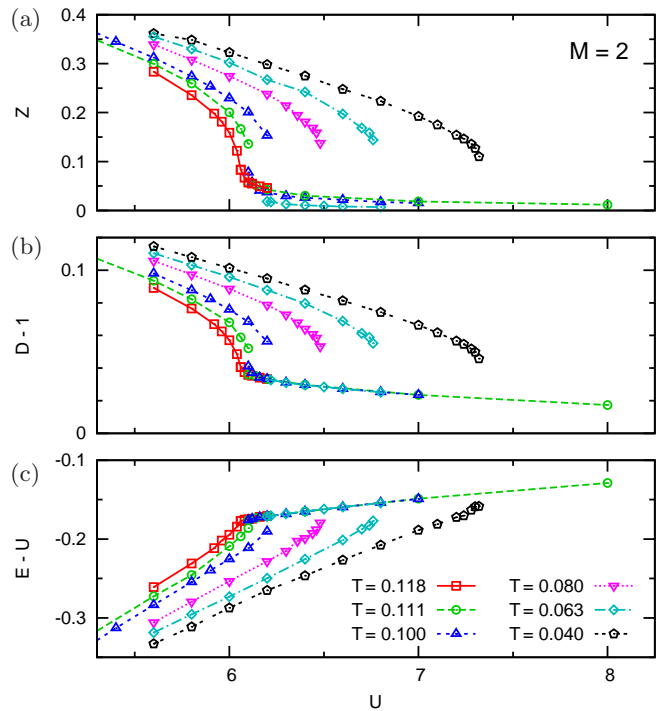}}
  \end{picture}
  \caption{(Color online)  Numerically exact DMFT+QMC results (symbols) for $M=2$: (a) quasiparticle weight $Z$, (b) pair occupancy $D$, shifted by its value in the atomic limit, (c) energy $E$, relative to the asymptotics in the atomic limit; lines are guides to the eye only.
\label{fig:Z_D_E_M2}}
\end{figure}
However, as more than two fermions can occupy the same site for $N>2$, it is advantageous to generalize the concept of the double occupancy to that of the pair occupancy $D=\sum_{\alpha<\alpha'}\langle n_{i\alpha} n_{i\alpha'}\rangle$; we have retained the symbol ``$D$'' for this observable, as it obviously reduces to the double occupancy in the single-band case and satisfies the relation $E_{\text{int}} = D U$, stated in the previous section, for arbitrary degeneracy $N$ (or number of orbitals $M$). At fixed integer band filling $n$, its minimum value as a function of $U$ and $T$ is $D_\mymin=n(n-1)/2$, corresponding to an atomic state with exactly $n$ filled orbitals;
while this minimum is zero in the single-band case at half filling ($n=1$), it has the values 1, 6, and 28 in the half-filled case ($n=M$) at $M=2$, $M=4$, and $M=8$, respectively.\cite{fn:uncorr}
For better comparison to the previous results, we have, therefore, subtracted $D_\mymin=1$ in \reffb{Z_D_E_M2}. The corresponding interaction energy $E_\mymin=U D_\mymin = U$ has also been subtracted from the energy in \reffc{Z_D_E_M2}; only with this adjustment does the slope $\partial E/\partial U$ approach zero in the limit of strong interaction (in the insulating phase).

With these adjustments, the data shown in \reff{Z_D_E_M2} for $M=2$ look remarkably similar to the single-band case at low temperatures; in addition to data for $T<T^*$, i.e., with coexistence, we have also included results for $T=0.118$, where the DMFT solution is unique for all interactions, corresponding to a continuous crossover curve (solid line). 

By reading off the phase boundaries from these numerically exact QMC data we can, for the first time, construct the coexistence phase diagram for $M=2$ in an unbiased way, down to the lowest QMC temperature $T=0.04$, as denoted by circles in \reff{phase_M2}.
\begin{figure}[t] 
  \includegraphics[width=\columnwidth]{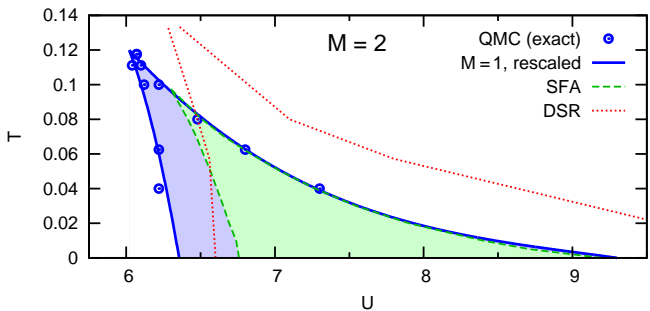}
  \caption{(Color online) Coexistence phase diagram for $M=2$ (within DMFT): exact QMC results (circles) for the boundaries at $T\ge 0.04$ are consistent with a fit (solid lines and shaded area) obtained by rescaling the exact $M=1$ coexistence region. Also shown: SFA result\cite{Inaba_PRB05} (dashed lines and green-shaded area) and DSR prediction\cite{Florens_PRB02a} (dotted lines). \label{fig:phase_M2}}
\end{figure}
Also shown is the SFA prediction\cite{Inaba_PRB05} (green dashed lines and green-shaded area) as well as the DSR result\cite{Florens_PRB02a} (dotted lines). Quite remarkably, the QMC estimates of $U_{c2}$ agree perfectly with the corresponding SFA prediction at $T<T^*_{\text{SFA}}$ (for $M=2$), even better than in the case $M=1$ (cf. inset of \reff{SFA_DSR}).
Consequently, we regard two-site SFA as practically exact (only) for $U_{c2}(T)$ at $M\gtrsim 2$ and will not try to compete with its estimates for the corresponding ground-state value $U_{c2}$. In contrast, the exact QMC results for $U_{c1}(T)$ are significantly lower than their SFA counterparts; also the QMC value for $T^*\approx 0.12$ is significantly above the SFA estimate. 
At the same time, the QMC data for $M=2$ (circles) are in excellent global agreement with a rescaled version (solid blue lines) of the numerically exact one-band result (solid blue lines in the inset of \reff{SFA_DSR}) determined above; thus, the coexistence regions for $M=1$ and $M=2$ appear to be similar even in the strict mathematical sense.

The DSR yield phase boundaries (dotted lines) of very similar shape, but shifted towards larger $U$ and $T$, with a prediction for $U^*$ which nearly coincides with that of the SFA. Our exact data now reveal that both estimates are too large by about 0.3. Still, both approximate methods, SFA and DSR, yield more accurate predictions of the phase diagram for $M=2$ than for $M=1$ (which in the case of the DSR might be due to a specific parameter choice); in particular, the discrepancies with respect to the area of the coexistence region (of about $20\%$) are much smaller.

The QMC results for $M=4$, shown in \reff{Z_D_E_M4}, include three temperatures very close to $T^*$: 
\begin{figure}[t] 
  \includegraphics[width=\columnwidth]{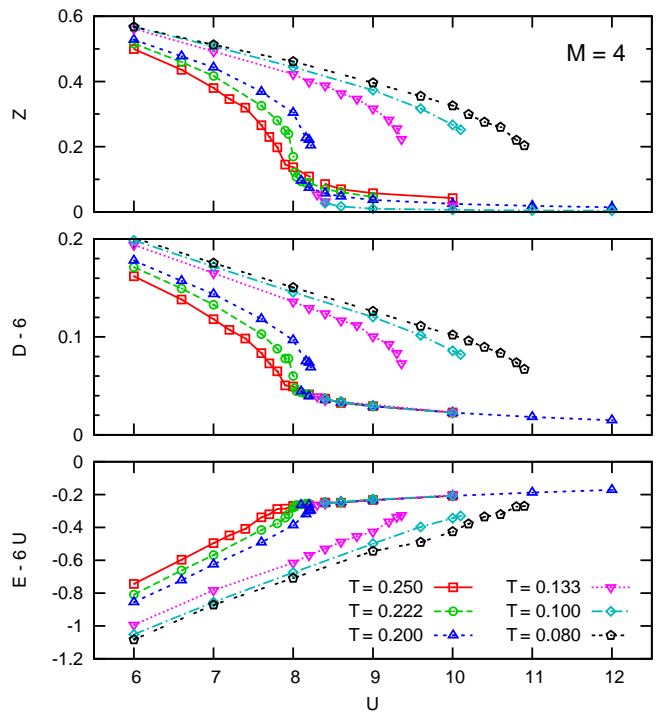}
  \caption{(Color online) Numerically exact DMFT+QMC results (symbols) for $M=4$, analogous to \reff{Z_D_E_M2}. \label{fig:Z_D_E_M4}}
\end{figure}
both at $T=0.25$ (squares) and at $T=0.222$ (circles) the curves are continuous, without coexistence, while a clear coexistence is observed at $T=0.2$ (upward triangles). At the same time, the maximum derivatives $\partial Z/\partial U$ and $\partial D/\partial U$ are much larger at $T=0.222$ than at the neighboring grid points; we conclude that $T^*\approx 0.22$. QMC results, which are already much more expensive computationally at $M=4$ than in the single-band case [the cost being roughly proportional to the number $M(2M-1)=28$ of Hubbard-Stratonovich fields] have also been obtained near $T^*/2$. Overall, the evolution of all three observables ($Z$, $D$, and $E$ as a function of $U$ and $T$) is consistent with the expectations from $M=1$ and $M=2$ (cf.\ \reff{Z_D_E_M1} and \reff{Z_D_E_M2}, respectively) within symbol sizes, which reflect approximate error bars.

Corresponding phase boundaries are shown as circles in \reff{phase_M4}.
\begin{figure}[t] 
  \includegraphics[width=\columnwidth]{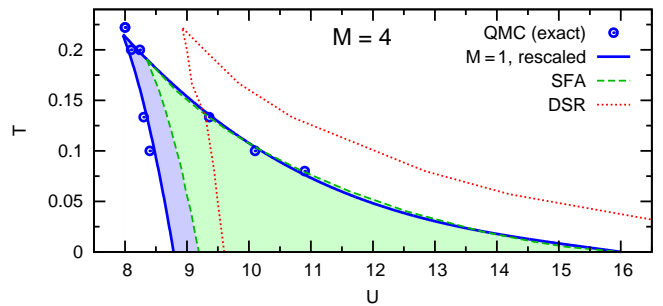}
  \caption{(Color online) Coexistence phase diagram for $M=4$ (within DMFT): exact QMC results (circles) and fit (solid lines and shaded area) in analogy with \reff{phase_M2}. Also shown: SFA result\cite{Inaba_PRB05} (dashed lines and green-shaded area) and DSR prediction\cite{Florens_PRB02a} (dotted lines). \label{fig:phase_M4}}
\end{figure}
These data confirm again, both the accuracy of the SFA prediction for $U_{c2}$ and the validity of the scaling assumption for the shape of the coexistence region, yielding the blue solid lines in \reff{phase_M4}.

The scaling assumption is also supported by the observed convergence of the SFA results for $U_{c1}(T)$ towards these rescaled one-band results (blue solid lines) in the series $M=1$ (inset of \reff{SFA_DSR}), $M=2$ (\reff{phase_M2}), and $M=4$ (\reff{phase_M4}); in the last case, the SFA discrepancy in $T^*$ has already shrunk to about $5\%$ and that in the area of the coexistence region to about $10\%$. While the DSR yields an essentially correct value of $T^*$ at $M=4$, the whole DSR coexistence region appears shifted towards larger interactions (relative to the unbiased QMC results, denoted by circles and solid lines) by an offset of roughly $1/8$ of its true width, similarly to the case $M=2$. We conclude that the DSR, in contrast to the SFA (with one bath site per orbital), does not become more accurate at large $M$.

Let us, finally, turn to the case of $M=8$ orbitals (i.e., a total degeneracy of 16 in the spin+orbital space), which has never been considered in the literature before. Due to the extreme computation cost (increased by a factor of 120 relative to the 1-band case), and since we have already established the universal shape of the phase diagram, we focus on temperatures in the immediate vicinity of the critical point. The QMC results depicted in \reff{Z_D_E_M8} show an increased scatter, indicating larger relative error bars as represented by the increased symbol sizes.
\begin{figure}[t] 
  \includegraphics[width=\columnwidth]{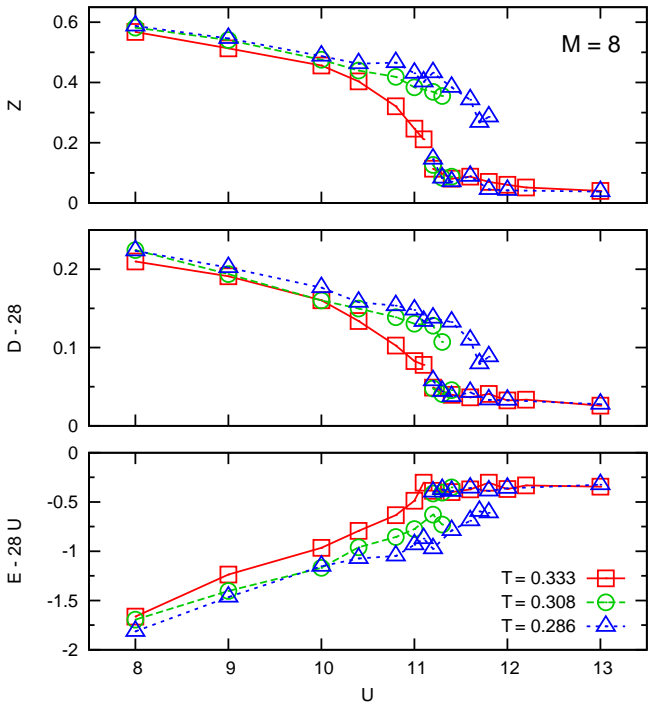}
  \caption{(Color online) Numerically exact DMFT+QMC results (symbols) for $M=8$, analogous to \reff{Z_D_E_M2} and \reff{Z_D_E_M4}. Increased symbol sizes reflect larger error bars. \label{fig:Z_D_E_M8}}
\end{figure}

Still, they allow us to locate the critical point at $T^*\approx 0.33$, $U^*\approx 10.9$. 
Together with the value $U_{c2}\approx 29.6$ read off from \reff{scaling}, these parameters also determine $U_{c1}\approx 12.8$ and (using the known shape) the full coexistence phase for $M=8$, shown in \reff{phase_M8}.
\begin{figure}[b] 
  \includegraphics[width=\columnwidth]{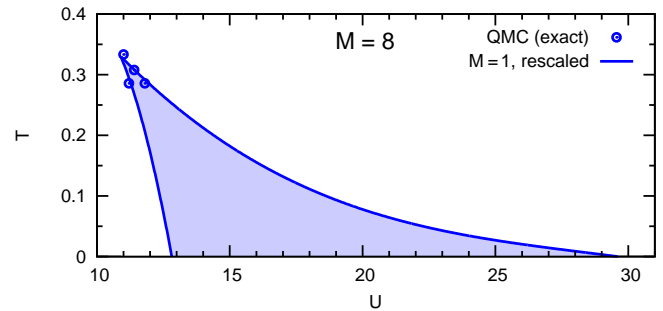}
  \caption{(Color online)  Coexistence phase diagram for $M=8$ (within DMFT): exact QMC results (circles) and fit (solid lines and shaded area) in analogy with \reff{phase_M2} and \reff{phase_M4}.\label{fig:phase_M8}}
\end{figure}
Due to the exponential scaling of exact diagonalization with the total number of orbitals (interacting and bath), an SFA solution analogous to those shown for $M=1,2,4$ would be prohibitively expensive at $M=8$.


\section{Scaling of critical parameters with spin and orbital degeneracy}\label{sec:scaling}
As seen in the preceding section, the critical parameters $T^*$, $U^*$ (of the finite-temperature critical end point) and $U_c$ (of the ground-state Mott transition) all increase significantly with increasing degeneracy [i.e., with the number $M$ of orbitals, corresponding to $N=2M$ fold degeneracy in the spin+orbital space]. In order to study the dependencies in detail, the left column of \reff{scaling} shows the estimates of these parameters as a function of the inverse number of orbitals, $M^{-1}$. At the scale of \reff{scaling} (b), all finite-temperature methods give quite similar results for $U^*$, with slight deviations for DSR (triangles and dotted line) at large degeneracy. Deviations become much more apparent for $T^*$, shown in \reff{scaling} (c), with SFA data (diamonds and dashed line) having a nearly constant negative offset relative to the exact QMC data (circles and solid line). Regarding $U_{c2}$, we see in \reff{scaling} (a) that DSR is far above the accurate SFA data at larger $M$; we have also included the L-DMFT prediction $U_{c2}=4M+2$. Obviously, all of the observables increase strongly towards smaller $1/M$ (i.e., towards larger degeneracy); alas, it is hard to distinguish exponents at this level.
\begin{figure}[t] 
\unitlength0.1\columnwidth
\begin{picture}(10,8)
   \put(0,0){\includegraphics[height=0.78\columnwidth]{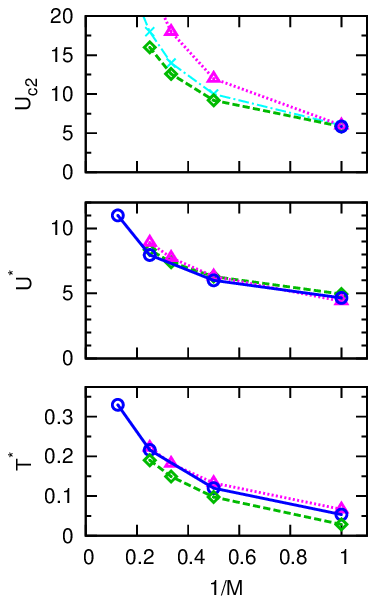}}
   \put(5.1,0){\includegraphics[height=0.78\columnwidth]{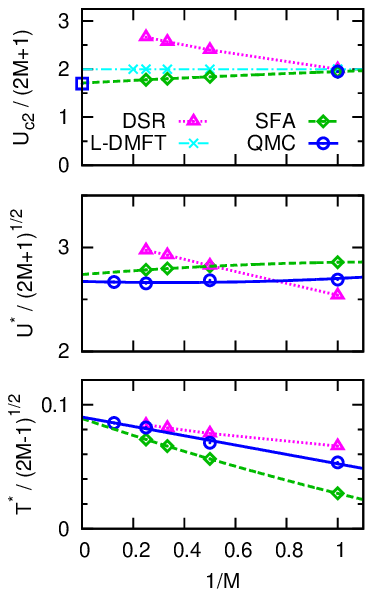}}
   \put(4.05,7.15){(a)}
   \put(4.05,4.75){(b)}
   \put(4.05,2.32){(c)}
   \put(9.25,7.15){(d)}
   \put(9.25,4.75){(e)}
   \put(9.25,2.32){(f)}
\end{picture}
  \caption{(Color online) Dependence of critical parameters on
band degeneracy. Left column: estimates of $U_{c2}$ (first row), $U^*$ (second row), and $T^*$ (third row) as a function of the inverse number of orbitals, $M^{-1}$, from DSR,\cite{Florens_PRB02a} linearized DMFT,\cite{Ono2003} SFA\cite{Imada1998} and numerically exact QMC calculations. Right column: rescaled critical parameters  are perfectly linear as a function of $M^{-1}$. The open square in the $U_{c2}$ scaling corresponds to the exact result for $U_{c2}$ at $M\to\infty$.\cite{Florens_PRB02b}
\label{fig:scaling}}
\end{figure}

The scaled data shown in the right hand column of \reff{scaling} demonstrate convincingly, however, that $U_{c2}$ indeed scales as $M$ while $U^*$ scales as $M^{1/2}$; in addition, we establish that also $T^*$ scales as $M^{1/2}$. In \reff{scaling} (d) we have, specifically, divided $U_{c2}$ by $2M+1$ (instead of $M$) in order to convert the L-DMFT prediction to a constant (with value 2). With this particular scaling ansatz, also the SFA data fall on a straight line, interpolating between the numerically exact results for $M=1$ and the analytic expression for $M=\infty$ (square). Our fit corresponds to the scaling law
\begin{equation}\label{eq:Uc2}
   U_{c2} \approx 1.70\, (2M+1)\,\big(1 + 0.166\, M^{-1}\big)\,.
\end{equation}
In this representation, DSR is even seen to have the wrong tendency; this method should be off by more than a factor of two for $M\to\infty$.

The same offset in the argument is also seen, in \reff{scaling} (e), to minimize curvature when rescaling estimates of $U^*$ (to $U^*/\sqrt{2M+1}$). Specifically, the exact QMC data become nearly flat and can safely be extrapolated to $1/M=0$, with the result (given without higher order corrections as they are not significant)
\begin{equation}\label{eq:Us}
  U^* \approx 2.67\, \sqrt{2M+1}\,.
\end{equation}
The SFA data are significantly above the QMC results at all finite $M$. In the extrapolation to $1/M=0$ some discrepancy remains; it is not entirely clear whether it is significant.

For rescaling $T^*$, we have chosen a different offset in \reff{scaling} (f) which ensures, again, that the results of each method fall on a nearly straight line, with perfect convergence of the SFA data to the exact QMC results. We conclude that $T^*$ is well represented by the expression
\begin{equation}\label{eq:Ts}
  T^* \approx 0.090\, \sqrt{2M-1}\, \big(1 - 0.41\, M^{-1}\big)\,.
\end{equation}
Note that both corrections to the asymptotic scaling $T^*\propto M^{1/2}$, arising from the shift in the argument and associated with the explicit $1/M$ term, work in the same direction: in the physical range of $M$, the critical temperature increases much faster with the degeneracy than one would expect from the scaling law. For example, going from SU(2) to the SU(6) Hubbard model, recently realized with ultracold rare-earth atoms,\cite{Taie2012} increases $T^*$ by a factor of $3.3$, much larger than the factor $\sqrt{3}\approx 1.73$ suggested by the large-$M$ asymptotics. This extra enhancement is certainly beneficial for accessing Mott physics in cold atom experiments.

Let us stress again, that all numerical results correspond to a semi-elliptic density of states of unit variance (and full bandwidth 4); they can be converted to the cubic lattice, in units of the hopping $t$, by multiplication with $\sqrt{6}\approx 2.45$ (or to the square lattice by multiplication with $\sqrt{4}=2$).\cite{fn:scales}

The expressions \refq{Uc2} -- \refq{Ts} fully determine the coexistence phase diagram at any orbital degeneracy $M$ when combined with the scaling phase diagram \reff{phase_scaled} in which, by construction, the finite-temperature critical point has the coordinates $(0,1)$ while the ground-state critical point has the coordinates $(1,0)$ for any value of $M$. 
\begin{figure}[t] 
  \includegraphics[width=\columnwidth]{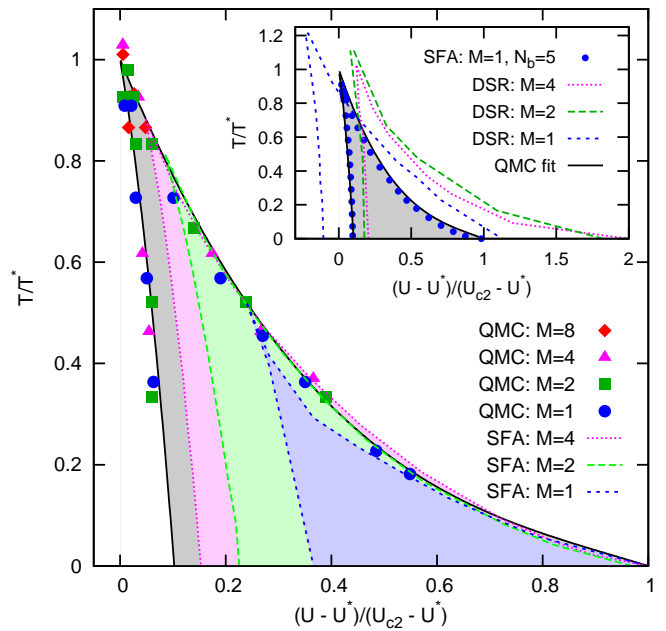}
  \caption{(Color online) Scaling phase diagram: upon rescaling with the parameters $T^*$, $U^*$ of the second-order finite-temperature critical end point and with the critical interaction $U_{c2}$ of the second-order ground-state Mott transition, the exact QMC data (symbols) collapse onto the scaling curves (solid lines). SFA results (broken lines and shaded regions) deviate, but converge towards this scaling form for large $M$. Inset: DSR results\cite{Florens_PRB02a} (broken lines) deviate from the scaling form (solid lines) at all $M$; in contrast, a high-precision SFA calculation\cite{Pozgajcic2004} (with 5 bath sites: circles) nearly recovers the reference result already at $M=1$.
\label{fig:phase_scaled}}
\end{figure}
Its main panel shows that the QMC data (symbols) for the phase boundaries indeed collapse onto a universal phase diagram (black solid lines and gray shaded background) in this representation, while the SFA data (using one bath site per interacting orbital) approach it only at large $M$. As seen in the inset of \reff{phase_scaled}, the inclusion of a larger number of bath sites (instead of one per orbital) vastly improves the accuracy of the SFA also at $M=1$ (filled circles),\cite{Pozgajcic2004} beyond the level of the multi-band case with the same total number of sites. The inset further shows that the DSR data seem to converge after rescaling (with a near-collapse between the results for $M=2$ and $M=4$), but to an incorrect limit.

From the universal phase diagram, one can also read off that the insulating state is meta-stable (within DMFT and in a paramagnetic phase) at zero temperature down to 
\begin{equation}\label{eq:Uc1}
  U_{c1} \approx 0.9\, U^* + 0.1\, U_{c2} \,;
\end{equation}
which is easily expressed explicitely in terms of $M$ by using Eqs.\ \refq{Uc2} and \refq{Us}.

\section{Conclusion and outlook}
  In conclusion, we have studied the Mott metal-insulator transition of the SU($2M$) symmetric Hubbard model by solving the paramagnetic DMFT equations numerically exactly. Our results confirm the predicted\cite{Florens_PRB02b} asymptotics of the ground-state critical interaction $U_{c2}\propto M$ for $M\to\infty$ and determine the (previously unknown) subleading corrections. They also confirm the predicted\cite{Koch1999} exponent (of $1/2$) of the dependence of the finite-temperature critical interaction $U^*$ on $M$ and yield the missing prefactor plus subleading terms; in addition, they establish the relation $T^*\propto M^{1/2}$ also for the associated critical temperature. Despite the different scaling of the end points of the first-order phase-transition line with $M$, the shape of the coexistence region is found to be universal to an astonishing degree. This universality could only be revealed by a method (multigrid HF-QMC) that is numerically exact at arbitrary $M$; in earlier SFA and DSR studies, it was obscured by systematic errors. 

Our results yield precise predictions for the Mott transition at arbitrary values of $M$, to be tested in cold-atom experiments. Due to the enhanced critical temperatures, the multi-flavor case might make Mott physics more accessible than in the single-band (i.e., two-flavor) case, in which the Mott signatures\cite{Joerdens2008,Schneider2008} seen so far correspond to crossovers, not true phase transitions. On the other hand, the experimental two-flavor studies profited from the fact that the MIT extends, as a crossover, far above $T^*$ with relatively little variation in $U$; thus, it is possible, e.g., to obtain good estimates of $U^*$ from  measurements at $T\gtrsim T^*$. This is still true in the SU(3) case.\cite{Gorelik2009} At large M, however, the relative variation of $U$ along the MIT line increases significantly, from $(U_{c2}-U^*)/U_{c2}\approx 0.2$ at $M=1$ to, e.g., $(U_{c2}-U^*)/U_{c2}\approx 0.6$ at $M=8$.\cite{fn:M_to_intfy_ratio}
One may suspect that the relative variation of $U$ in the crossover region is similarly enhanced at large degeneracy, which implies that a closer approach of $T^*$ would be required in order to determine $U^*$. Such low-temperature experiments might also explore ordering phenomena, which are a fascinating topic of their own and beyond the scope of this paper.

More generally, our results provide high-precision numerical benchmarks for evaluating DMFT impurity solvers in the challenging regime of moderate to high orbital degeneracy; they could also be used for assessing the relative importance of nonlocal correlations at higher band degeneracy, e.g. by comparison with high-temperature expansions\cite{Hazzard2012} or with direct exact calculations once they become available.

  Support by the
  DFG within the Collaborative Research Centre SFB/TR 49 and by the John von Neumann Institute for Computing is
  gratefully acknowledged.




\begin{thebibliography}{16}
\bibitem{fn:scales} The different shapes of the densities of states (including different band edges and van Hove singularities) lead to slight corrections of these generic scaling factors. In particular, the noninteracting ground-state energy is $E_0=-2t$ for the cubic lattice which implies that $U_{c2}\to 8t$ for $M\to\infty$, i.e., the asymptotic critical value is larger by a factor of $3\pi/4\approx 2.36$ compared to the Bethe lattice, not by the generic factor $\sqrt{6}\approx 2.45$.
\bibitem{fn:uncorr} In the uncorrelated limit $U\to 0$, the pair occupancy approaches a value of $D_0=2M(2M-1)/8$ at half filling, i.e., 0.25, 1.5, 7, and 30 for $M=1$, $M=2$, $M=4$, and $M=8$, respectively. This implies that the dynamic range of $D$ scales linearly with the band degeneracy: $D_0-D_\mymin = M/4$.
\bibitem{fn:M_to_intfy_ratio} For $M\to\infty$, this ratio approaches unity; then, the ordinate in the scaling phase diagram \reff{phase_scaled} can be replaced by $U/U_{c2}$. This means that, in units of $U_{c2}$, the finite-temperature critical interaction $U^*$ vanishes.
\end{thebibliography}
\end{document}